\documentclass[12pt,preprint]{aastex}

\newcommand{\sbr}{ergs cm$^{-2}$ s$^{-1}$ sr$^{-1}$}
\newcommand{\kms}{km s$^{-1}$}
\newcommand{\cms}{cm$^{-3}$}
\newcommand{\feii}{[\ion{Fe}{2}]}
\newcommand{\hh}{H$_2$}

\slugcomment{Submitted: \today}

\shorttitle{Near-Infrared Observations of 3C~396}
\shortauthors{Lee et al.}

\begin{document}

\title{Near-Infrared \feii\ and \hh\ Line Observations of 
the Supernova Remnant 3C~396: 
Probing the Pre-supernova Circumstellar Materials}

\author{Ho-Gyu Lee\altaffilmark{1,2}, Dae-Sik Moon\altaffilmark{3},
Bon-Chul Koo\altaffilmark{1}, Jae-Joon Lee\altaffilmark{4},
and Keith Matthews\altaffilmark{5}}

\altaffiltext{1}{Department of Physics and Astronomy, 
Seoul National University, Seoul, 151-742, Korea; 
hglee@astro.snu.ac.kr, koo@astrohi.snu.ac.kr}
\altaffiltext{2}{Astrophysical Research Center 
for the Structure and Evolution of the Cosmos, 
Sejong University, Seoul 143-747, Korea}
\altaffiltext{3}{Department of Astronomy and Astrophysics, 
University of Toronto, Toronto, ON M5S 3H4, Canada; 
moon@astro.utoronto.ca}
\altaffiltext{4}{Department of Astronomy and Astrophysics, 
Pennsylvania State University, University Park, PA 16802; 
lee@astro.psu.edu}
\altaffiltext{5}{Division of Physics, Mathematics and Astronomy, 
California Institute of Technology, Pasadena, CA 91125; 
kym@caltech.edu}

\begin{abstract}
We present the results of near-infrared \feii\ and \hh\ line imaging
and spectroscopic observations of the supernova remnant 3C~396
using the Palomar 5~m Hale telescope.
We detect long, filamentary \feii\ emission delineating the inner edge
of the radio emission in the western boundary of the remnant in imaging observations,
together with a bright \feii\ emission clump close to the remnant center.
There appears to be faint, diffuse \feii\ emission between the central
clump and the western filamentary emission.
The spectroscopic observations determine the expansion velocity of the central clump
to be $\sim 56$ \kms. This is far smaller than the expansion velocity of 3C~396
obtained from X-ray observations, implying the inhomogeneity of the ambient medium.
The electron number density of the \feii\ emission gas is $\le$ 2,000 \cms.
The \hh\ line emission, on the other hand,
lies slightly outside the filamentary \feii\ emission in the
western boundary, and forms a rather straight filament.
We suggest that the \feii\ emission represents dense clumps in the wind material
from the red supergiant phase of a Type IIL/b progenitor of 3C~396 which
have been swept up by the supernova remnant shocks.
The \hh\ emission may represent either the boundary of a wind bubble produced during
the main-sequence phase of the progenitor or molecular clumps left over inside the bubble.
We propose that the near-infrared \feii\ and \hh\ emission observed in several 
supernova remnants of Type IIL/b SNe likely has the same origin.
\end{abstract}


\section{Introduction}

The near-infrared (IR) \feii\ lines are useful tools for studying
the radiative shocks of supernova remnants (SNRs).
The increased abundance of the gas phase iron by
shock-induced sputtering of the dust grains and/or
the creation of an extensive partially ionized zone by shock heating
can substantially enhance the \feii\ emission \citep[e.g.,][]{gre91, mou00}.
This, together with less severe extinction effects than the optical
and the recent advents of near-IR imaging cameras covering sufficiently wide fields
(e.g., $\ge$ 5\arcmin),
can potentially make the observations of the near-IR \feii\ lines
a very competitive and efficient way to study the radiative shocks of SNRs,
especially in the Galactic plane where the extinction is usually large.

A good example is the young core-collapse SNR G11.2$-$0.3 where
we recently found the brightest near-IR \feii\ emission of all
known SNRs using Wide-field Infrared Camera aboard the
Palomar 5-m telescope \citep{koo07}.
Besides G11.2$-$0.3, a few other SNRs --
such as RCW~103 \citep{oli99}, 3C~391 \citep{rea02}, W44 \citep{rea05},
and W49B \citep{keo07} --
have been detected in the near-IR \feii\ line
emission.\footnote{Here we only consider \feii\ emission from the radiative SNR shocks,
excluding \feii\ emission from the supernova ejecta \citep[e.g.,][]{moo08}}
Note that all these SNRs have a core-collapse nature
and, except for W44, all of them are relatively young SNRs.
Interestingly, in addition to the \feii\ lines,
near-IR \hh\ emission of shocked molecular gas
has been detected from all these SNRs.
The \hh\ emission is distributed close to, but slightly outside, the \feii\ emission
from the center of the SNRs \citep[e.g.,][]{oli90, keo07, koo07},
which is opposite to the standard picture of the molecular shocks
where molecules form in the downstream behind a recombination region.
This apparent pattern of the positional `reversal' between
the \feii\ and \hh\ emission
may have important clues to understanding the pre-supernova environment of SNRs,
calling more systematic investigation of the \feii\ and \hh\ emission together
in (young) core-collapse SNRs.

\begin{figure}
\plotone{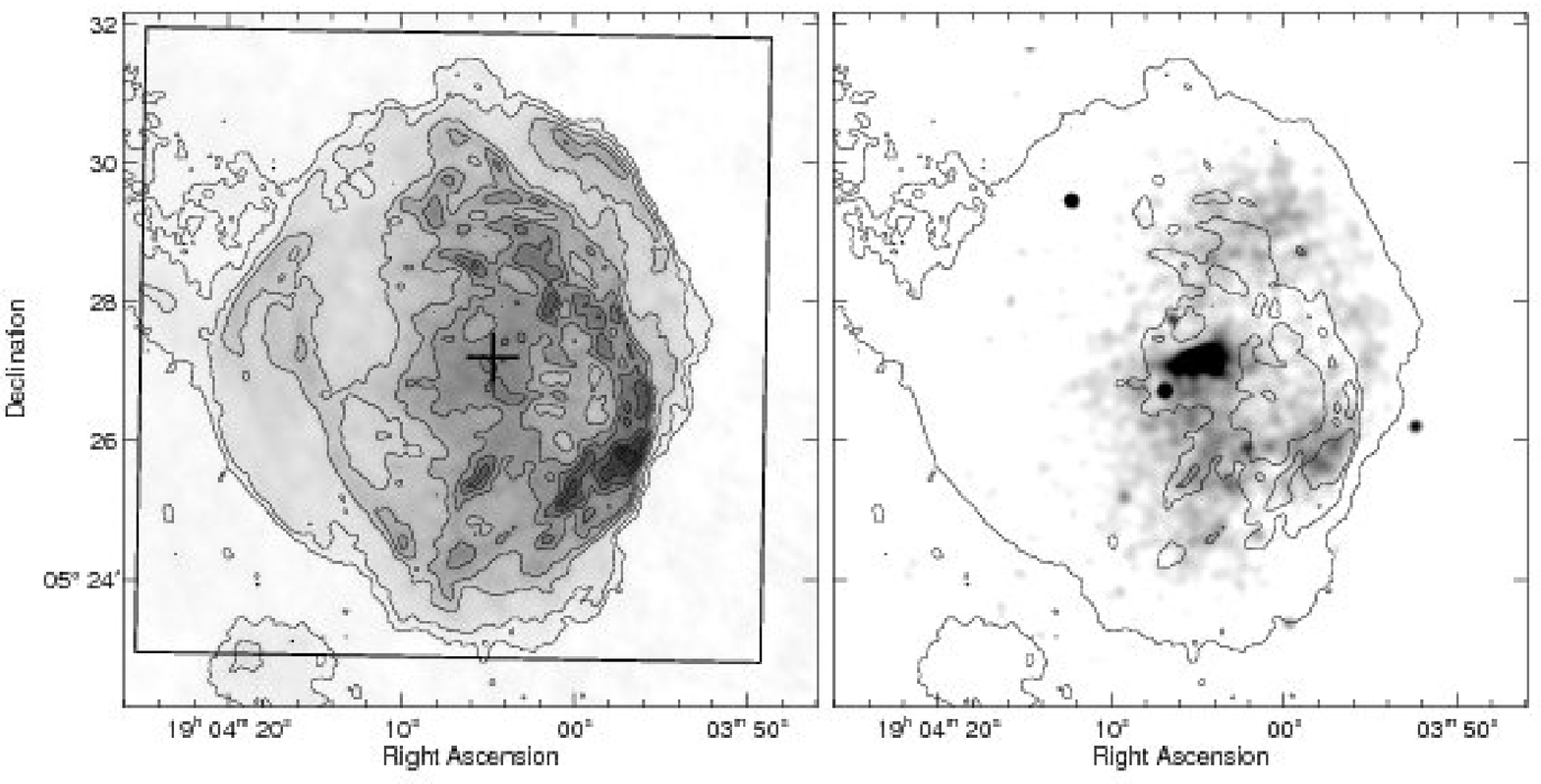}
\caption{
$(Left)$ VLA 20~cm radio continuum image of 3C~396 \citep{dye99}.
The contour levels are 1, 2, 4, 8, 10, 12, 14, and 16 mJy beam$^{-1}$,
and the beam size is $6''.8 \times 6''.1$.
The large box inside indicates the field covered by our near-IR imaging observations with WIRC.
The cross at the center marks the central position of its X-ray pulsar wind nebula.
$(Right)$ Chandra X-ray image of 3C~396 \citep{olb03} superimposed on
radio contours of 1, 8, 12 mJy beam$^{-1}$.
}
\label{fig_rx}
\end{figure}

In this paper we present the near-IR imaging and spectroscopic
observations of \feii\ and \hh\ lines from another core-collapse
SNR 3C~396 (also known as G39.2--0.3).
In radio 3C~396 is a shell-type SNR of $\sim$ $8'\times6'$ size \citep{and93}
composed of multiple elliptical arcs and filaments (Figure~\ref{fig_rx}),
with a thick, half ring-like feature in the west.
The {\em ASCA} X-ray observation revealed
the existence of a pulsar wind nebula within
a thermal X-ray shell \citep{har99},
and the former was later resolved by {\em Chandra},
although no X-ray pulsation was detected \citep{olb03}.
Recently {\em Spitzer} detected bright mid-IR emission
from this SNR along its western boundary in the radio \citep{lee05, rea06}.
The mid-IR emission is filamentary,
and is prominent in the 4.5 and 5.8 $\mu$m bands.
If the origin of the mid-IR emission is ionic and molecular lines
as suggested \citep{lee05, rea06},
we expect the existence of the corresponding near-IR
emission of the \feii\ line at 1.64 $\mu$m and \hh\ line at 2.12 $\mu$m
from this SNR, which motivated our observations presented in this paper.
The distance to 3C~396 is rather uncertain; however,
most of the previous observations of H and OH indicate that it is
beyond the tangential point at 6.6~kpc,
and possibly at 9.6~kpc \citep{cas75, bec87, gre89}.
In this paper, we adopt the distance of 8.5 kpc determined based on
the velocity of a nearby molecular cloud in 3C~396 (see \S~4).
We organize this paper as follows.
We first give a description of our near-IR imaging and spectroscopic
observations of 3C~396 in \S~2,
together with the basic data reduction processes.
In \S~3 and \S~4,
we present the results of the \feii\ and \hh\ line observations,
respectively.
We discuss the origin of the \feii\ and \hh\ emission in \S~5
and summarize the paper in \S~6.

\section{Observations and Data Reduction}

Near-IR narrow-band imaging observations of the SNR 3C~396
were carried out to search for \feii\ and \hh\ emission
using Wide-field Infrared Camera (WIRC)
aboard the Palomar 5-m Hale telescope on 2005 July 15 and August 28 and 29.
WIRC is equipped with a 
Rockwell Scientific now Teledyne Hawaii II HgCdTe 2K
IR focal plane array, covering a $\sim$ 8\farcm5 $\times$ 8\farcm5
field of view with a 0\farcs25 pixel scale.
Our WIRC field covers the entire area of 3C~396 seen in radio
except for a tiny portion of the faint tail in the east
(see Figure~\ref{fig_rx}).
The narrow-band filters for the 1.64~$\mu$m \feii\ line and 2.12~$\mu$m \hh\
line were used, together with $H$-cont and $K$-cont filters to subtract
out continuum emission.
The detailed observing parameters are given in Table~\ref{tab_obsimg}.
For the basic data reduction, the dark and sky background were subtracted
out from each dithered frame using the median filtering of
stacked dithered frames, and then
the dithered frames were combined to a final image after flat fielding.
The photometric solutions of the \feii\ and \hh\ images were
obtained using the aperture photometry of
14 isolated stars from the 2MASS point source catalog \citep{skr06}.
The obtained instrumental magnitudes of the \feii\ and \hh\
images were found to have a very good correlation ($r \simeq 1$) with the
2MASS $H$ and $K_{\rm s}$ magnitudes.
Astrometric solutions were also obtained using the 2MASS stars.
The 1-$\sigma$ uncertainties of both R.A. and decl. directions
in the \feii\ and \hh\ images were smaller than 0\farcs05
which is within the systematic accuracy of the 2MASS astrometry.

\begin{figure}
\plotone{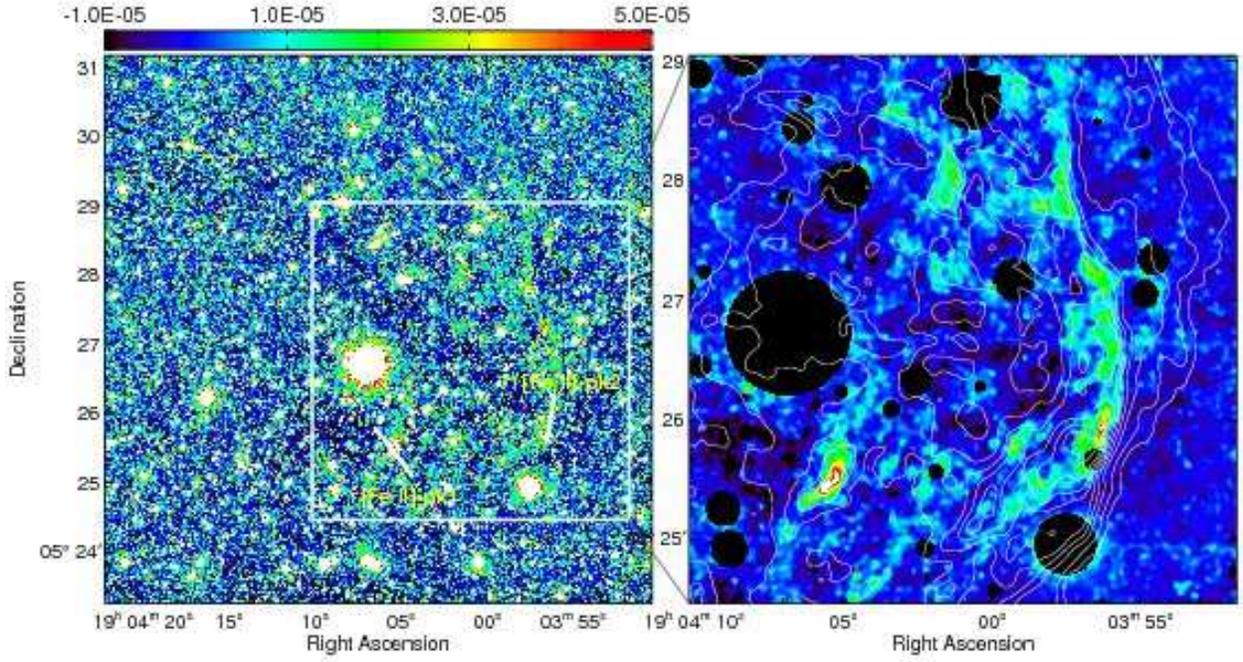}
\caption{
$(Left)$ WIRC image of 3C~396 obtained with
the \feii\ 1.64 $\mu$m narrow-band filter.
The filamentary \feii\ emission is detected in the western part of remnant
distinguished from the point-like stellar emission.
The two slit positions used for the spectroscopic observations
are indicated by elongated white bars in the small internal panel.
The surface brightness scale range of two panel images is expressed 
by the color-bar at the top in unit of \sbr.
$(Right)$ Enlarged image of the panel in the left after the subtraction of
stellar emission, superimposed on radio contours.
Median-box filtering and Gaussian smoothing are applied
to enhance the image quality (see \S~2).
}
\label{fig_fe}
\end{figure}

\begin{figure}
\plotone{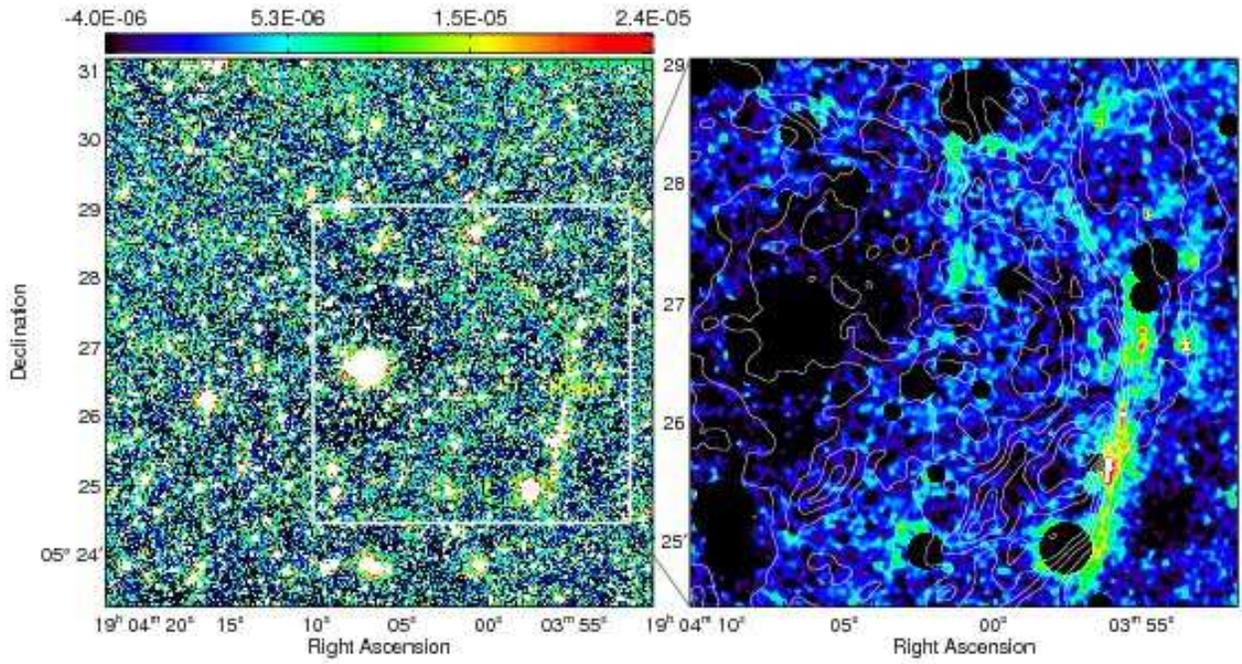}
\caption{
Same as Figure~\ref{fig_fe} but for the \hh\ emission.
}
\label{fig_h2}
\end{figure}

The left panels in Figures~\ref{fig_fe} and \ref{fig_h2} present
the WIRC images obtained with
the \feii\ 1.64 $\mu$m and \hh\ 2.12 $\mu$m narrow-band filters respectively,
where we see filamentary non-stellar emission in both the \feii\ and \hh\ images
which are highly contaminated by the emission from numerous stars in the field.
In order to see the non-stellar emission more clearly
we subtracted out the stellar emission using the continuum
emission images obtained with the $H$-cont and $K$-cont filters.
For this we first performed PSF photometry of the $H$-cont image
to obtain a list of stars for continuum subtraction,
and then subtracted out the corresponding stars in the \feii\ image.
To enhance the image quality we applied a median filtering with
a square box of 1\farcs75 $\times$ 1\farcs75  and then smoothed the resulting
image using a Gaussian with an 1$''$ FWHM.
It was difficult to remove several very bright stars with the above method,
so we simply masked them out.
We followed the same procedure with the \hh\ and $K$-cont images.
The right panels in Figures~\ref{fig_fe} and \ref{fig_h2} present
the star-subtracted images obtained with the \feii\ and \hh\ filters,
and we can clearly identify the non-stellar extended emission in 3C~396.
We give detailed analysis of the \feii\ and \hh\ emission in \S~3 and 4.

After the aforementioned imaging observations,
follow-up spectroscopic observations were carried out using the
Long-slit Near-IR Spectrograph aboard the same Hale Telescope \citep{lar96}.
The spectrograph is equipped with a 256 $\times$ 256 HgCdTe NICMOS detector
from Rockwell International now Teledyne,
providing a 38\arcsec\ slit length
and 1\arcsec\ (= 6 detector pixels) slit width.
Both low-resolution ($R$ $\simeq$ 700) and high-resolution ($R$ $\simeq$ 5,000)
modes of the spectrograph were used for the observations.
The spectral coverages of the low-resolution mode were
$\sim$ 0.06 and 0.12 $\mu$m
for the \feii\ (1.64 $\mu$m) and \hh\ (2.12 $\mu$m) lines respectively,
while that of the high-resolution mode was
$\sim$ 0.02 $\mu$m for the \feii\ line.
(Note that only the \feii\ line was observed with the high-resolution mode.)
The \feii\ line spectra were obtained toward two bright peaks
of the 1.64 $\mu$m \feii\ line emission found in the WIRC image (see \S~3).
The coordinates (J2000) of these two peaks,
which are referred to \feii-pk1\ and \feii-pk2\ in this paper,
are (R.A., decl.) =
($\rm 19^h04^m05.54^s$, $\rm +05^\circ25'26.8''$)
and ($\rm 19^h03^m56.29^s$, $\rm +05^\circ25'56.0''$) respectively.
As in Figure~\ref{fig_fe} the \feii-pk1\ is located inside 3C~396 while
\feii-pk2\ is at its western boundary.
The \hh\ line spectrum was obtained toward the peak position at
($\rm 19^h03^m56.08^s$, $\rm +05^\circ25'38.3''$),
which is referred to \hh-pk1.
The parameters of the spectroscopic observations are listed
in Table~\ref{tab_obssp}.
Both flat fielding and atmospheric opacity correction were
performed by dividing the spectrum of a standard G-type star
obtained at a similar airmass just after the target observations.
Then the target spectrum was multiplied by a blackbody curve with
an effective temperature of the standard G-type star.
The OH air glow lines \citep{rou00} were used for wavelength calibration
and the Earth motion was corrected.

\section{Results of \feii\ Observations}

\subsection{Distribution of \feii\ Emission in 3C~396}
 
We can clearly identify extended, non-stellar \feii\ emission
spread over the central and the western parts of 3C~396 in Figure~\ref{fig_fe}.
The brightest \feii\ emission is from the
\feii-pk1 knot which is about $\sim$ 1\arcmin\ (or 2 pc) south from the center.
Its surface brightness is $7.6 \pm 0.5 \times 10^{-5}$ \sbr,
and the $\sim$~30\arcsec-long (or 1 pc) northwest-southeast elongation
matches well the local radio emission.
In addition to the emission around the \feii-pk1 knot,
there is a long, curved, half ring-like \feii\ filament in the
western boundary of 3C~396.
This filament, which peaks at \feii-pk2,
is clumpy and $\sim$ 3\arcmin\ (or 7 pc) long.
Its peak and average surface brightnesses are
$4.6 \pm 0.5 \times 10^{-5}$ and $1.5 \pm 0.5 \times 10^{-5}$ \sbr, respectively.
Compared to the radio emission,
this filamentary \feii\ emission is confined by radio contours
delineates the inner edge of a large, bright half-shell feature
prominent in radio in the western part of the remnant
(Figures~\ref{fig_rx} and \ref{fig_fe}).
The radio half-shell feature is relatively thick ($\sim$ 0\farcs5\ or 1 pc),
and the brightnesses of the both \feii\ and radio emission drop
sharply across the western boundary of this feature.
The central \feii\ emission and the emission in the west
appear to be connected though low-surface brightness \feii\ emission.
There is also more diffuse \feii\ emission in the north around
(R.A., decl.) = ($\rm 19^h03^m58^s$, $\rm +05^\circ29'10''$),
although it might be an extension of the filamentary emission in the west.
The mean surface brightness of the emission in the north is
$0.8 \pm 0.5 \times 10^{-5}$ \sbr.
Integrating all the mentioned features above,
the total flux of the \feii\ 1.64 $\mu$m emission in 3C~396 is
$1.8 \pm 0.6 \times 10^{-12}$ erg cm$^{-2}$ s$^{-1}$.

\subsection{\feii\ Spectroscopic Observations}

\subsubsection{Low-Resolution Spectroscopy}

\begin{figure}
\plotone{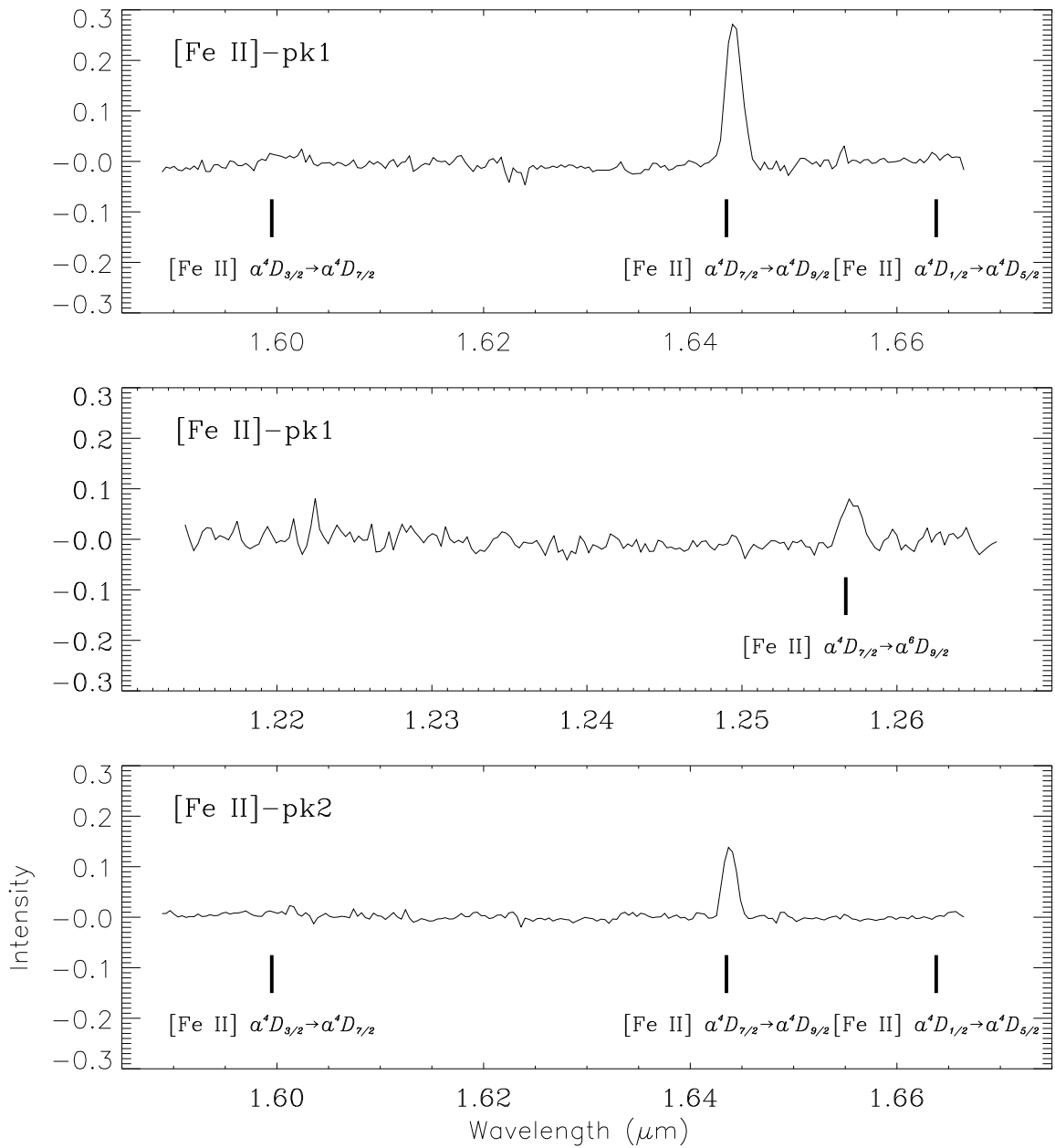}
\caption{
$(Top)$ Low-resolution spectrum of the \feii\ 1.64 $\mu$m line transition from \feii-pk1.
The air wavelengths of the \feii\ transitions are indicated.
$(Middle)$ Same as the top panel, but for the \feii\ 1.26 $\mu$m line transition.
$(Bottom)$ Same as the top panel, but for \feii-pk2.
}
\label{fig_splow}
\end{figure}

Figure~\ref{fig_splow} presents low-resolution spectra of
\feii-pk1 and \feii-pk2.
For \feii-pk1, we placed the slit perpendicular to the elongation
direction of the clump, while the slit direction was parallel to
the direction of the filamentary emission for \feii-pk2 (see Figure~\ref{fig_fe}).
We have averaged extended emission over $\sim$ 5\arcsec\ (for \feii-pk1)
and 27\arcsec\ (for \feii-pk2) regions around their peak emission to obtain the spectra.
The near-IR \feii\ emission is known to have several other line transitions
besides the transition at 1.64 $\mu$m, including the transitions
at 1.26, 1.60, and 1.66 $\mu$m \citep[e.g.,][]{koo07}.
We clearly detected the 1.64 $\mu$m transition
from both \feii-pk1 and \feii-pk2, while we detected the 1.26 $\mu$m transition
only from \feii-pk1.
(Note that our \feii-pk2 spectrum does not cover the 1.26 $\mu$m transition.)
For all other transitions we obtain the upper limits
of their intensities relative to that of the 1.64 $\mu$m transition.
Table~\ref{tab_splow} summarizes the results.

The \feii\ 1.26 and 1.64 $\mu$m transitions originate
from the same upper level.
Thus their intensity ratio can provide information
on the extinction toward the source, given the Einstein $A$ coefficients.
The exact values of the $A$ coefficients, however, are somewhat controversial
and their ratio could be from 1.04 \citep{qui96} to 1.49 \citep{smi06}.
From \feii-pk1, we obtain \feii\ 1.26 to 1.64 $\mu$m line ratio
of $0.232 \pm 0.026$,
which gives the visual extinction in the range of $A_{\rm V}$ = 17 -- 21 mag
based on the extinction cross section per hydrogen nuclei for the
carbonaceous-silicate model
of interstellar dust with $R_{\rm V}$ = 3.1 \citep{dra03}.
The estimated column density of hydrogen nuclei
is $N_H$ = 3.4 -- 4.2 $\times$ 10$^{22}$ cm$^{-2}$,
which is between the values obtained by X-ray and
radio H~I observations \citep{bec87, har99, olb03}.
We adopt $N_{\rm H}$ = 3.4 $\times$ 10$^{22}$ cm$^{-2}$,
which is approximately the middle value of X-ray and H I observations,
in this paper.
On the other hand,
the intensity ratios of the two undetected \feii\ transitions
at 1.60 and 1.66 $\mu$m to the 1.64 $\mu$m transition is useful to estimate
the upper limit of electron number density \citep[e.g.,][]{koo07}.
We solve the rate equation of 16 levels of the ionized iron using the
atomic parameters assembled by the CLOUDY program
\citep[ver. C05.05;][]{fer98},
and derived an upper limit of the electron number density to be
$\la$~2,000~cm$^{-3}$ for both \feii-pk1 and \feii-pk2
at assumed temperature of 5,000~K.

\subsubsection{High-Resolution Spectroscopy}

\begin{figure}
\plotone{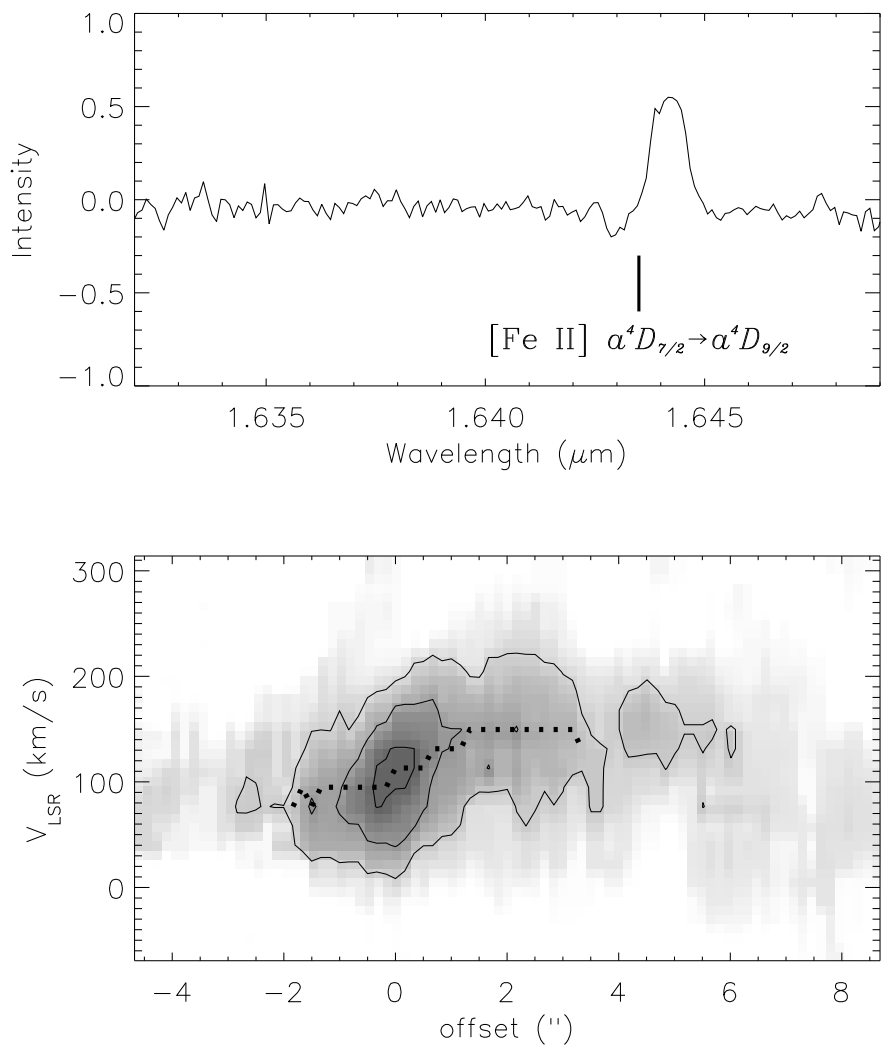}
\caption{
$(Top)$ High-resolution spectrum of the \feii\ 1.64 $\mu$m line transition from \feii-pk1.
Note that the center of the line profile is shifted from the rest wavelength of the transition
indicated by the vertical line, clearly showing a Doppler shift.
$(Bottom)$ Position-velocity diagram of the high-resolution spectrum in the top panel.
The dotted line represents the evolution of the velocity centroids along the offset.
The negative offset corresponds to the northeast along the slit.
}
\label{fig_sphi}
\end{figure}

Figure~\ref{fig_sphi} (top) is the high-resolution spectrum of
\feii\ 1.64 $\mu$m emission from \feii-pk1
averaged over a 10$''$ area around the peak.
The central velocity and the FWHM of the line
are $v_{LSR}$ = +117 $\pm$ 18 \kms\ and 149 $\pm$ 5 \kms, respectively.
(For comparison, the velocity of \feii-pk2 in the western boundary
measured from the low-resolution spectrum in Figure~\ref{fig_fe}
is +50 $\pm$ 26 \kms, which is close to
the systematic velocity +69 \kms\ of 3C~396 caused by the Galactic rotation [see \S~4.1].)
The large linewidth of the averaged spectrum is partly due to a velocity shift
which is apparent in the position-velocity diagram in Figure~\ref{fig_sphi} (bottom)
where we can identify a velocity shift from 93 to 145 \kms\
across the peak position from the northeast to the southwest direction.
The measured velocity +117 \kms\ of \feii-pk1 corresponds to a radial
velocity of +48 \kms\ after subtraction of the contribution from the Galactic rotation.

If all the \feii\ emission in 3C~396 is originated from the surface
of an expanding sphere of 2\arcmin5 (or 6 pc) radius
which is the average distance of the western filament from the center,
then the three dimensional expansion velocity of \feii-pk1 is $\sim$ 56 \kms.
(Note that the radial velocity of the western filament is comparable to the
systematic Galactic rotational velocity of 3C~396,
indicating that the filament motion is largely transverse.)
The obtained expansion velocity of 
56 \kms\ is too low to be ejecta \citep[e.g.,][]{moo08},
which suggests that the \feii\ gas in 3C~396
is primarily the ambient or circumstellar gas swept up
by the SNR shock.
Generally the SNR shocks become radiative at expansion
velocities of $<$ 200 \kms\ \citep{mck80},
so that the above expansion velocity of 56 \kms\ of 3C~396
appears to be consistent with the detection of the shocked \feii\ lines
in this remnant.
In addition, the de-reddened surface brightness
(0.1 -- 1.2 $\times 10^{-3}$ \sbr) can be explained by
a shock propagating into the gas
of $n_{\rm H}$ = 10$^3$ -- 10$^4$~cm$^{-3}$
with a velocity of 30 -- 50 \kms\ \citep{har04}.

\section{Results of \hh\ Observations}

\subsection{Distribution of \hh\ Emission and Molecular Gas Toward 3C~396}
 
\begin{figure}
\plotone{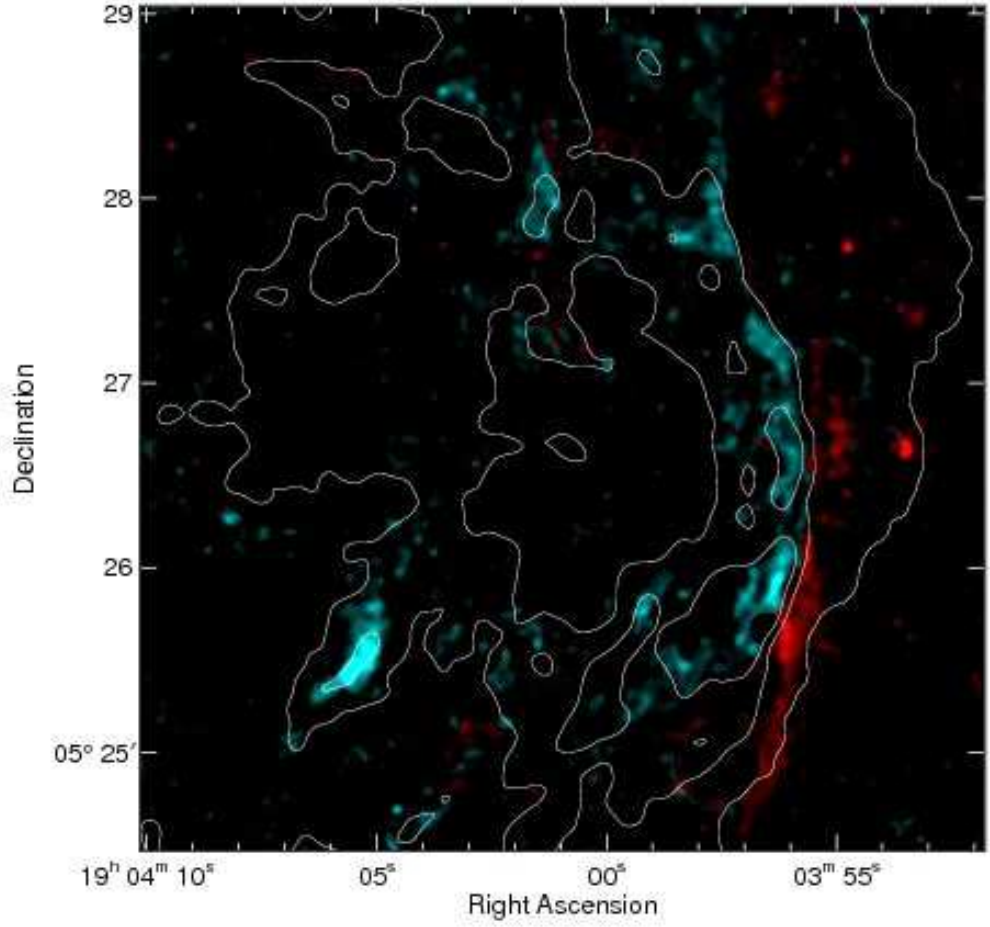}
\caption
{Comparison between the star-subtracted images of
the \feii\ and \hh\ emission superimposed on the radio contours in the western boundary of 3C~396.
Blue is for the \feii\ emission; red is for the \hh\ emission.
Note that the \hh\ emission is clearly located outside the \feii\ emission from the center.
}
\label{fig_twocolor}
\end{figure}

There is only one conspicuous feature in
the \hh\ 2.12 $\mu$m emission which is a long
($\sim$ 3\arcmin, or 7 pc)
filamentary emission along the southwestern boundary of 3C~396
(Figure~\ref{fig_h2}).
The peak position (\hh-pk1)
is slightly south of the position of \feii-pk2,
and its surface brightness is $2.4 \pm 0.2 \times 10^{-5}$ \sbr.
The total integrated flux of the \hh\ 2.12 $\mu$m emission is
$4.1 \pm 1.3 \times 10^{-13}$ erg cm$^{-2}$ s$^{-1}$.
Figure~\ref{fig_twocolor} compares the relative locations
of the \feii\ and \hh\ emission, superimposed on the radio contours.
It is apparent that the \hh\ emission is located slightly farther
than the \feii\ emission from the SNR center,
and the former is linear as opposed to the latter which is curved
along the bright radio emission.
The \feii\ and \hh\ filaments are adjacent near the position of \feii-pk1,
but are apparently separated at the edges.

\begin{figure}
\plotone{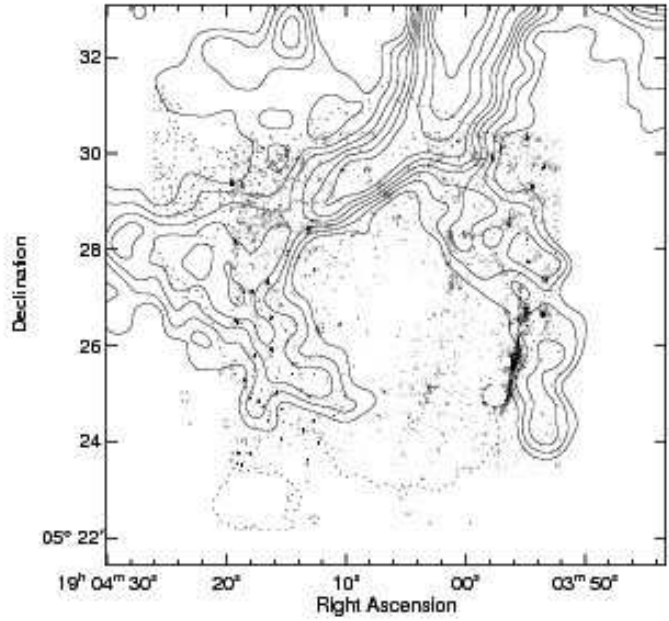}
\caption
{Comparison between the WIRC \hh\ 2.12 $\mu$m image of 3C~396 and
contours of the $^{13}$CO J=1--0 transition integrated
over the velocity range of 68 -- 70 \kms (see text).
The radio-continuum boundary of the remnant is marked by dotted line.
}
\label{fig_co}
\end{figure}

Across the southwest boundary of 3C~396,
the radio emission drops sharply (Figure~\ref{fig_rx}).
This, together with our detection of the \hh\ emission,
strongly suggests that 3C~396 has encountered dense medium in this region,
most likely a molecular cloud.
In order to investigate this more
we examined the distribution of molecular gas around 3C~396
using the Galactic Ring $^{13}$CO J=1--0 line Survey \citep{jac06}
and found molecular gas at $v_{\rm LSR}$ = 68--70 \kms\ 
surrounding 3C~396 (Figure~\ref{fig_co}).
These velocities are consistent with the previously proposed systematic velocity
of 3C~396 \citep{cas75},
and the location of the \hh\ filament that we found in this study
matches the inner boundary of the molecular gas that forms a ring-like structure.
Therefore, it is likely that 3C~396 is indeed interacting with
some of these molecular clouds.
If we adopt the velocity (69 \kms) of the molecular gas 
as the systemic velocity of 3C~396,
its distance is 8.5 kpc which we use in this paper as the distance to 3C~396.

\subsection{Spectroscopic Observations and Molecular Shocks}

\begin{figure}
\plotone{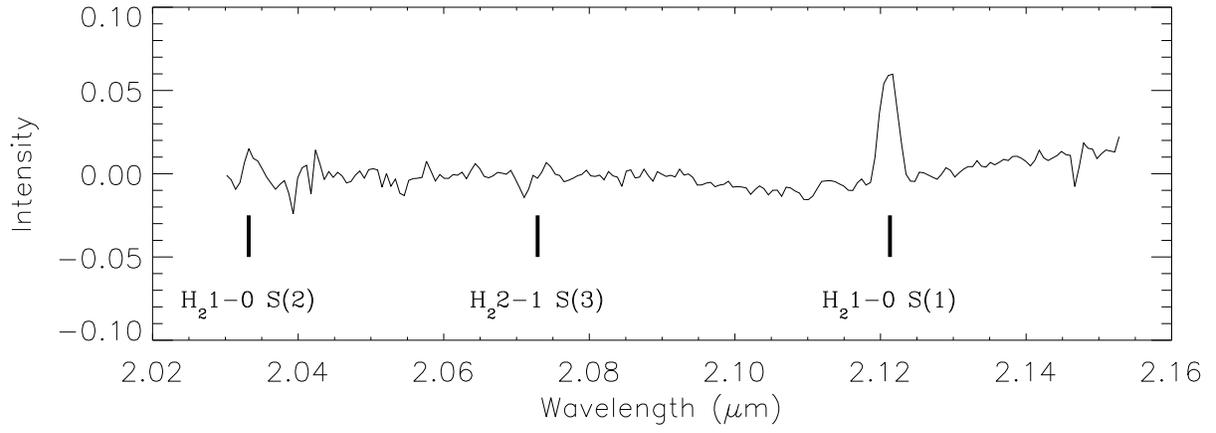}
\caption{Low-resolution spectrum of \hh\ 2.12 $\mu$m transition from \hh-pk1.
The air wavelengths of the \hh\ transitions are indicated.
}
\label{fig_sph2}
\end{figure}

Our \hh\ spectrum of \hh-pk1 averaged over a 24$''$ area around the peak 
is shown in Figure~\ref{fig_sph2}.
The \hh\ 2.12 $\mu$m 1-0 S(1) line stands out clearly,
while the detection of \hh\ 2.03 $\mu$m 1-0 S(2) line is marginal.
The \hh\ 2.07 $\mu$m 2-1 S(3) line is not detected,
which could be partly due to the incomplete OH airglow subtraction.
The ratio of the \hh\ 2.03 $\mu$m 1-0 S(2) line to the 2.12 $\mu$m 1-0 S(1) 
line is $0.25~\pm~0.08$,
giving an excitation temperature of $T_{ex}$ = $820~\pm~350$~K
based on the transition probabilities of \citet{wol98} 
and an ortho to para hydrogen ratio of 3.
This is lower than the general value of $\sim 2,000$ K derived from
the ro-vibrational \hh\ 1-0 S(1) and 2-1 S(1) line ratios in
thermalized molecular shock regions \citep{bur89}.
It may be that in the case of 3C~396 the density of the molecular gas is
smaller than the critical density for thermalization,
so that the excitation temperature is lower.

The dereddened surface brightness of \hh-pk1 is
$1.4~\times~10^{-4}$ \sbr.
This surface brightness can be produced by a C-shock
penetrating into a molecular cloud of 
n$_H$ = $10^{4}$ cm$^{-3}$ and $B$ = 50 -- 100 $\mu$G
with a shock velocity of 20--30 km s$^{-1}$ \citep{dra83}.
Since we are observing tangentially to the shock front in 3C~396,
the real density could be lower than the model value 
which was calculated for a normal surface brightness.

\section{Origin of the \feii\ and \hh\  Filaments}

We detected the \feii\ 1.64 $\mu$m and \hh\ 2.12 $\mu$m filaments
which are close to each other in the western boundary of 3C~396.
While the \feii\ filament is curved and distributed 
along the bright radio emission,
the \hh\ filament appears to be straight and is located just 
{\it outside} the bright radio emission.
The \feii\ and \hh\ filaments are adjacent in the middle,
but are apparently separated at the edges.
The former is brighter than the latter.

What would be the origin of these filaments?
In principle, one can observe the emission from both Fe$^+$ ions and \hh\ molecules
in a single SNR shock if the shock is radiative and has swept up enough column density
to have reformation of molecules behind the shock \citep[e.g.,][]{hol89b}.
However, in such cases the \hh\ emission originates in the downstream,
which is opposite to the `reversal' of the relative positions between the \feii\ and \hh\
emission observed in 3C~396 (Figure~\ref{fig_twocolor}).
Interestingly 3C~396 is not the first SNR
where the `reversal' is detected.
It has been detected previously in several SNRs, including
RCW~103 \citep{oli89, oli90}, G11.2$-$0.3 \citep{koo07}, and W49B \citep{keo07}.
(The `reversal' was also observed in W44 \citep{rea05},
but it is excluded in our discussion because it is an old SNR
and also \feii\ emission is fainter than \hh\ emission,
on the contrary to the other SNRs.)
Table~\ref{tab_snrlist} summarizes the properties of these four SNRs (including 3C~396)
where we list their distances, radii, and ages, along with
their interactions with ambient molecular clouds.
For the origin of the `reversal' between the near-IR \feii\ and \hh\ emission,
\citet{oli90} considered the excitation of \hh\ by ultraviolet or X-rays from a SNR,
although it appears unlikely because of the difficulty in exciting the cold gas.
For W49B, \citet{keo07} proposed an explosion inside a wind bubble within a molecular cloud and
\hh\ emission being produced by a `jet' propagating beyond the bubble; however,
it is inadequate to explain all the four SNRs.

All the SNRs in Table~\ref{tab_snrlist}
are core-collapse SNRs with a relatively small radius of $\la 10$ pc.
According to \cite{che05}, G11.2$-$0.3 and RCW~103 are remnants of Type IIL/b SNe
that had intermediate-mass (25--35 $M_\odot$) progenitors with strong red supergiant (RSG) winds,
and the SNR shocks are interacting with the wind material extending to 5--7~pc.
The radio morphology of 3C~396, together with the detection of the \feii\ and \hh\ filaments,
indicates that it may also be a remnant of a Type IIL/b SN.
In the radio emission (Figure~\ref{fig_rx}),
there exists a thick, half ring-like bright radio shell (BRS) in the western part of the remnant.
Beyond this BRS of $\sim$ 2\farcm5 (or 6 pc) radius,
there appears faint plateau-like emission especially in the north and south of the remnant.
We interpret this BRS as the RSG wind material confined by the ambient pressure
and the faint extended plateau-like emission as originating from the SNR shock
that has crossed the boundary of the wind material.
This interpretation is consistent with the size of the BRS which is comparable to
to the typical size of the RSG wind material confined by the ambient pressure \citep{che05}.
In this scenario, the \feii\ emission may trace the shocked material of the RSG wind,
while the \hh\ emission may represent interactions between the SNR shock and
a wind blown bubble produced during the main sequence (MS) phase of the progenitor star
by its fast stellar winds \citep[e.g.,][]{mck84b}.
This also provides a natural explanation of the positional `reversal' between
the \feii\ and \hh\ emission.
The \hh\ filament could be either the inner boundary of the MS wind blown bubble
or molecular clumps left over inside the bubble.
We note that the 4.5 and 5.8 $\mu$m emission of 3C~396 detected with $Spitzer$
\citep[][; see \S~1]{lee05, rea06} can also be explained by this scenario.
The $Spitzer$ 4.5~$\mu$m waveband contains the wavelengths of
pure rotational \hh\ lines, while the 5.8~$\mu$m one does those
of many atomic lines, including the \feii\ line at 5.34~$\mu$m.
The 4.5~$\mu$m emission is located slightly outside of the 5.8~$\mu$m emission,
consistent with what we see in this study with the near-IR \hh\ and \feii\ line emission.
In conclusion, the two-stage SNR shock interactions with RSG wind material and also with
a MS wind blown bubble appear to explain the near-IR \feii\ and \hh\ emission
in the young SNRs of Type IIL/b SNe, including 3C~396.

We suggest that the \feii\ emission in 3C~396 represent dense clumps
in the RSG wind of the progenitor star as follows.
The expansion velocity of \feii-pk1
is very small compared to what we expect from a young SNR.
If we use the hot gas temperature of 0.62 keV
derived from X-ray observations \citep{har99},
the implied velocity of the SNR shock in 3C~396 is in fact 600~\kms.
This discrepancy indicates that the ambient medium may be inhomogeneous,
so that the SNR shocks propagating into dense clumps are slow and radiative
while the shocks propagating through the diffuse interclump medium
are fast and non-radiative.
Generally in a shocked gas the \feii\ lines are emitted in the far downstream
where hydrogen atoms are mostly neutral at $T = 10^3$--$10^4$~K 
\citep{mck84a, hol89a, oli89}.
This is because the ionization potential of the Fe atom is low (7.9 eV)
and the far-ultraviolet photons from hot shocked gas can penetrate far downstream
to maintain the ionization state of Fe$^+$.
The upper limit of the electron density that we derived in \S~3.2.1
gives an upper limit of the density of hydrogen nuclei
$n_H \approx n_e/0.11\lesssim 2\times 10^4$~cm$^{-3}$
if the mean ionization fraction is 0.11 \citep{oli89}.
Assuming 5,000~K as the temperature of the \feii\ line emitting region,
the thermal pressure is $p/k_B\lesssim 1\times 10^8$~cm$^{-3}$ K where
$k_B$ is Boltzmann constant, which is comparable to the thermal pressure of
the X-ray emitting hot gas \citep{har99}.
This indicates that the shock in the \feii\ gas can be driven
by the thermal pressure of the hot gas.
The preshock density may be estimated from
$n_0\approx p/{\mu_H v_s^2}\sim 50$~cm$^{-3}$ where
$\mu_{\rm H}$ (= $2.34\times 10^{-24}$ g) is the mean mass
per hydrogen nuclei including the cosmic abundance of He and $v_s\simeq 56$~\kms\
is the shock speed.
The density of the RSG wind at $r=6$ pc is $n_w=\dot M/4\pi r^2 \mu_H v_w\sim 1$~cm$^{-3}$
if we use the mass loss rate $\dot M=5\times 10^{-5}$~$M_\odot$ yr$^{-1}$
and the wind speed $v_w=15$~\kms\ \citep[e.g.,][]{che05}.
Therefore, the preshock density of the \feii\ emitting gas
is an order of magnitude greater than the mean density of the RSG wind,
suggesting that the former represents dense clumps in the RSG wind material.

\section{Conclusion}

Stars of intermediate-mass (25--35 M$_\odot$) have fast winds
in the MS phase and slow dense winds during the RSG phase.
Just before a core-collapse SN explosion, therefore,
they are surrounded by a dense circumsteller region of 5--7~pc radius
created by the RSG winds embedded within a low-density bubble
produced by the MS winds \citep{che05}.
Once explodes, the SN shock encounters these pre-existing structures
to generate in principle a multi-shell structure \citep[e.g.][]{dwa05},
although the details can differ from case to case depending
on the distribution of the interstellar medium.
In this paper, based on the detection of the near-IR \feii\ and \hh\ filaments,
we propose that the SNR 3C~396 is a remnant of a Type IIL/b SN that shows
such a multi-shell structure produced by the interactions of the SNR
shocks with the RSG wind material (i.e., \feii\ filament)
and also with the MS wind blown bubble (i.e., \hh\ filament).
This scenario can provide a natural explanation of the relative locations
between the \feii\ and \hh\ emission, where the latter is placed farther
from the center of the remnant than the former, and is also consistent
with the $Spitzer$ 4.5 and 5.8 $\mu$m emission of 3C~396.
The \feii\ emission likely represents dense clumps driven by the thermal
pressure of hot X-ray emitting gas in the remnant.
We identify that there exists a molecular cloud that appears to surround 3C~396.
The velocities of the molecular gas are very similar to the systematic
velocity of 3C~396, indicating that this molecular gas may be indeed interacting
with 3C~396 to produce the \hh\ emission.
Further sensitive molecular line observations are needed to investigate this
possibility more.
Finally, we note that there are three other SNRs having \feii\ and \hh\ emission
somewhat similar to that of 3C~396.
We expect that these SNRs are also remnants of Type IIL/b SNe \citep[e.g.,][]{che05}
that have comparable environments produced by the stellar winds of similar progenitors.

\acknowledgments
We would like to thank Kristy Dyer for providing the VLA image of 3C~396.
This work was partly supported by Korea Science and Engineering Foundation
through the Joint Research Project under the KOSEF-NSERC Cooperative Program
(F01-2007-000-10048-0).
D.-S.M. acknowledges the support from the Discovery Grant (327277) of
Natural Science and Engineering Research Council of Canada.

{\it Facility:} \facility{Hale (WIRC, HNA)}

\clearpage

\clearpage

\clearpage

\clearpage

\clearpage

\clearpage

\clearpage

\clearpage

\clearpage
\begin{deluxetable}{lccccc}
\tabletypesize{\scriptsize}
\tablewidth{0pt}
\tablecolumns{5}
\tablecaption{Parameters of WIRC Imaging Observations \label{tab_obsimg}}
\tablehead {
\colhead{Filter} & \colhead{$\lambda _c\tablenotemark{a}$} 
   & \colhead{$\Delta \lambda \tablenotemark{b}$ }
   & \colhead{Exposure$\tablenotemark{c}$} &  \colhead{Date} }
\startdata
$[$Fe II$]$& 1.644 $\mu$m & 0.0252 $\mu$m & 900 s & 2005. 07. 15. \\
H$_2$    & 2.120 $\mu$m & 0.0329 $\mu$m & 600 s & 2005. 08. 28. \\
$H$-cont & 1.570 $\mu$m & 0.0236 $\mu$m & 300 s & 2005. 08. 29. \\
$K$-cont & 2.270 $\mu$m & 0.0330 $\mu$m & 300 s & 2005. 08. 29. \\
\enddata
\tablenotetext{a}{The central wavelength of the filter obtained from 
$\rm http://www.astro.caltech.edu/palomar/200inch/wirc/wirc\_spec.html$}
\tablenotetext{b}{Equivalent width of
$\Delta \lambda~=~\int S(\lambda) d\lambda$ 
where $S(\lambda)$ is the normalized filter response. }
\tablenotetext{c}{The integrated exposure over dithered frames of 30~s exposure.}
\end{deluxetable}

\begin{deluxetable}{lcccc}
\tabletypesize{\scriptsize}
\tablewidth{0pt}
\tablecolumns{5}
\tablecaption{Parameters of Spectroscopic Observations \label{tab_obssp}}
\tablehead {
\colhead{Position} 
   & \colhead{Central wavelength}
   & \colhead{Exposure} & \colhead{Mode$\tablenotemark{a}$} &  \colhead{Date} }
\startdata
$[$Fe II$]$-pk1  & 1.24 $\mu$m &  600 s & Low  & 2005. 08. 26. \\
                      & 1.63 $\mu$m &  600 s & Low  & 2005. 08. 26. \\
                      & 1.64 $\mu$m & 1200 s & High & 2006. 06. 09. \\
$[$Fe II$]$-pk2  & 1.63 $\mu$m &  300 s & Low  & 2005. 08. 26. \\
H$_2$-pk1        & 2.09 $\mu$m &  300 s & Low  & 2006. 06. 10. \\
\enddata
\tablenotetext{a}{~Low: Low-resolution mode; High: High-resolution mode.}
\end{deluxetable}

\begin{deluxetable}{lclll}
\tabletypesize{\scriptsize}
\tablewidth{0pt}
\tablecolumns{5}
\tablecaption{Relative Intensities of the \feii\ and \hh\ Lines \label{tab_splow}}
\tablehead{
\colhead{Position} & \colhead{Wavelength}   & \colhead{Transition}    
&\multicolumn{2}{c}{Relative Strength} \\
                   &                        &                         
&\colhead{Observed}& \colhead{Dereddened} }
\startdata
\underline{[Fe II]-pk1}
   & 1.26 $\mu$m  & [Fe II] $a^4D_{7/2} \rightarrow a^6D_{9/2}$ 
       & 0.232 (0.026)       & 1.05 (0.13) \\ 
   & 1.60 $\mu$m  & [Fe II] $a^4D_{3/2} \rightarrow a^4F_{7/2}$
       & $<$ 0.034           & $<$ 0.039   \\
   & 1.64 $\mu$m  & [Fe II] $a^4D_{7/2} \rightarrow a^4F_{9/2}$
       & 1                   & 1           \\
   & 1.66 $\mu$m  & [Fe II] $a^4D_{1/2} \rightarrow a^4F_{5/2}$
       & $<$ 0.028           & $<$ 0.027   \\
\underline{[Fe II]-pk2}
   & 1.60 $\mu$m  & [Fe II] $a^4D_{3/2} \rightarrow a^4F_{7/2}$
       & $<$ 0.019           & $<$ 0.022   \\
   & 1.64 $\mu$m  & [Fe II] $a^4D_{7/2} \rightarrow a^4F_{9/2}$
       & 0.400 (0.014)    & 0.400 (0.014)  \\
   & 1.66 $\mu$m  & [Fe II] $a^4D_{1/2} \rightarrow a^4F_{5/2}$
       & $<$ 0.010           & $<$ 0.009   \\
\underline{H$_2$-pk}
   & 2.03 $\mu$m  & H$_2$ 1-0 S(2)
       & 0.247 (0.081)        & 0.284 (0.093) \\
   & 2.07 $\mu$m  & H$_2$ 2-1 S(3)
       & $<$ 0.023          & $<$ 0.025       \\
   & 2.12 $\mu$m  & H$_2$ 1-0 S(1)
       & 1                  & 1               \\
\enddata
\end{deluxetable}

\begin{deluxetable}{lcccccl}
\tabletypesize{\scriptsize}
\tablewidth{0pt}
\tablecolumns{7}
\tablecaption{Properties of SNRs with \feii\ inside \hh\ emission \label{tab_snrlist}}
\tablehead{
\colhead{Name}   & \colhead{Distance}   & \colhead{Radius}   & \colhead{Age}
&\colhead{SN Type}   & \colhead{MC\tablenotemark{a}}  
& References \\
                 & \colhead{(kpc)}      & \colhead{(pc)}     & \colhead{(yrs)}
&                &
}
\startdata
G11.2-0.3   &5           &3.3     &1600
            &IIL/b       &No      & \citet{koo07}
\\
RCW~103      &3.8         &5.0     &2000
            &IIL/b       &Yes     & \citet{oli90, oli99, par06}
\\

W49B        &11.4        &7       &4000
            &?           &No      & \citet{keo07}
\\
3C~396       &8.5         &10      &6000
            &IIL/b       &Yes     & This work
\\
\enddata
\tablenotetext{a}{Detection of an ambient molecular cloud 
interacting with the SNR}
\end{deluxetable}

\end{document}